\documentclass[prl,nofootinbib,floats,aps,superscriptaddress,twocolumn,preprintnumbers]{revtex4}
\usepackage{graphicx}
\usepackage{epsfig}
\begin{document}
%\draft
\newcommand{\be}{\begin{equation}}
\newcommand{\ee}{\end{equation}}
\newcommand{\bea}{\begin{eqnarray}}
\newcommand{\eea}{\end{eqnarray}}
\def\ie{{\it i.e.}}
%%%%%%%%%%%%%%%%%%%%%%%%%%%%%%%%%%%%%%%%%%%%%%%%%%%%%%%%%%%%%%%%%%%%%%%
\def\lsim{\raise0.3ex\hbox{$\;<$\kern-0.75em\raise-1.1ex\hbox{$\sim\;$}}}
\def\gsim{\raise0.3ex\hbox{$\;>$\kern-0.75em\raise-1.1ex\hbox{$\sim\;$}}}
\def\Frac#1#2{\frac{\displaystyle{#1}}{\displaystyle{#2}}}
\def\no{\nonumber\\}
\date{\today}
%%%%%%%%%%%%%%%%%%%%%%%%%%%%%%%%%%%%%%%%%%%%%%%%%%%%%%%%%%%%%%%%%%%%%%%
%
%\topmargin      -1.5cm  % distance to headers
\textwidth       180mm  % Horizontal alignment
\textheight      250mm  % height of text
\renewcommand{\baselinestretch}{1.12} \normalsize

\title{Supersymmetric contribution to the CP asymmetry of $B\to
J/\psi \phi$ in the light of recent $B_s -\bar{B}_s$ measurements}

\author{Shaaban Khalil}

\address{Ain Shams
University, Faculty of Science, Cairo, 11566, Egypt.\\
German University in Cairo-GUC, New Cairo city, Egypt.\\}
%\date{\today}
%\thispagestyle{empty}
\begin{abstract}
We derive new model independent constraints on the supersymmetric
extensions of the standard model from the new experimental
measurements of $B_s -\bar{B}_s$ mass difference. We point out
that supersymmetry can still give a significant contribution to
the CP asymmetry of $B_s \to J/\psi \phi$ that can be measured at
the LHCb experiment. These new constraints on the LL and RR squark
mixing severely restricted their possible contributions to the CP
asymmetries of $B\to \phi K$ and $B \to \eta' K$. Therefore, SUSY
models with dominant LR flavor mixing is the only way to
accommodate the apparent deviation of CP asymmetries from those
expected in the standard model. Finally we present an example of
SUSY non-minimal flavor model that can accommodate the new $\Delta
M_{B_s}$ results and also induces significant CP asymmetries in
$B_s \to J/\psi \phi$, $B\to \phi K$ and $B \to \eta' K$
processes.

\end{abstract}

\maketitle

%%%%
$1-$ Recently, the $D\emptyset$ \cite{Abazov:2006dm} and CDF
\cite{CDF} collaborations have reported new results for the $B_s
-\bar{B}_s$ mass difference: \bea 17~ ps^{-1} &<& \Delta M_{B_s} <
21~ ps^{-1}
~~~~ 90\%~C.L. ~~~(D\emptyset), \nonumber\\
\Delta M_{B_s} &=& 17.33^{+0.42}_{-0.21} \pm 0.07 ~ ps^{-1} ~~~
~~~~~~(CDF), \label{CDF}\eea which seems consistent with the
Standard Model (SM) predictions. In fact, the estimation of the SM
value for $\Delta M_{B_s}$ contains large hadronic uncertainties.
The $B^0_s-\bar{B}^0_s$ mass difference is defined as $\Delta
M_{B_s} = 2 {\cal M}_{12}(B_s) = 2 \vert \langle B^0_s \vert
H_{\rm{eff}}^{\Delta B=2} \vert \bar{B}^0_s\rangle \vert$, where
$H_{\rm{eff}}^{\Delta B=2}$ is the effective Hamiltonian responsible
for the $\Delta B=2$ transition. In the SM, $H_{\rm{eff}}^{\Delta
B=2}$ is generated by the box diagrams with $W$-exchange. The best
determination for $\Delta M^{\rm{SM}}_{B_s}$ can be obtained from a
ratio to the $\Delta M^{\rm{SM}}_{B_d}$ in which some QCD
corrections as well as $t$ quark mass dependence are cancelled out
\begin{equation} \frac{\Delta M^{\rm{SM}}_{B_s}}{\Delta
M^{\rm{SM}}_{B_d}}=\frac{M_{B_s}}{M_{B_d}}
\frac{B_{B_s}f^2_{B_{s}}}{B_{B_d}f^2_{B_{d}}}\frac{|V_{ts}|^2}{|V_{td}|^2},
\label{ratio} \end{equation} where $M_{B_d}=5.28$ GeV and
$M_{B_s}=5.37$ GeV and the lattice calculations lead to
$B_{B_s}f^2_{B_{s}}/(B_{B_d}f^2_{B_{d}})=(1.15\pm 0.06
^{+0.07}_{-0.00})^2$ \cite{lat1}. Since the $B^0_d-\bar{B}^0_d$
oscillation is mostly saturated by the SM contributions
\cite{Gabrielli:2002fr}, we can assume that $\Delta
M^{\rm{SM}}_{B_d}=\Delta M_{B_d}^{\rm{exp}} = (0.502\pm 0.007)
ps^{-1}$. Finally, $|V_{ts}|^2/|V_{td}|^2$ can be given as a
function of the angle $\gamma$ of the unitary triangle of
Cabibbo-Kobayashi-Maskawa (CKM) mixing matrix. In Fig. 1, we
present the allowed range of $\Delta M^{\rm{SM}}_{B_s}$ in terms
of the angle $\gamma$ (measured from a pure SM process). Here we
assume that $\vert V_{cb}\vert$ and $\vert V_{ub}\vert$ are free
of new physics and can be determined by the SM contribution to the
semileptonic decay. Also, it is assumed that the angle $\beta$ is
given by $\beta^{SM}$, measured from $B_d \to J/\psi K_s$. As can
be seen from this figure, the new bounds on $\Delta M_{B_s}$
impose stringent constraints on the values of $\gamma^{\rm{SM}}$.
The lower bound of $D\emptyset$ result excludes values of
$\gamma^{\rm{SM}} > 70^{\circ}$. It is worth mentioning that the
best fit for $\gamma^{\rm{SM}}$ and $\Delta M^{\rm{SM}}_{B_s}$,
according to UTfit group is given by \cite{UTfit}: \be
\gamma^{\rm{SM}}=61.3\pm 4.5, \Delta M^{\rm{SM}}_{B_s}=(17.45\pm
0.25) ps^{-1},\ee and according to CKMfitter group is given by
\cite{CKMfitter} \be \gamma^{\rm{SM}}=59.8^{+4.9}_{-4.1},~ \Delta
M^{\rm{SM}}_{B_s}=17.3^{+0.49}_{-0.20}. \ee

%
%----------------------------------------------
\begin{figure}[t]
\begin{center}
\epsfig{file=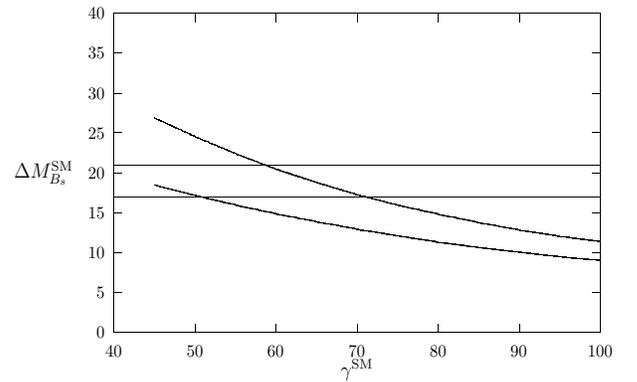, width=8 cm, height=5 cm, angle=0}
\end{center}
\caption{Allowed region of $\Delta M_{B_s}$, in the SM, as function
of the angle $\gamma$.} \label{DMSM}
\end{figure}
%-------------------------------------------------

Therefore, it is expected that the experimental measurements in
Eq.(\ref{CDF}) provide important constraints on any new physics
beyond the SM \cite{Ball:2006xx}. In this letter, we study the
constraints imposed on supersymmetric model due to these
experimental limits. We derive model independent bounds on the
relevant SUSY mass insertions. Then we analyze the implications of
these constraints on the supersymmetric contribution to the CP
asymmetry in $B_s \to J/\psi \phi$ process. Finally, we consider
the SUSY non-minimal flavor model studied in
Ref.\cite{Khalil:2005dx}, as an example for SUSY model, that can
accommodate the new $\Delta M_{B_s}$ results and also induces
significant CP asymmetry in $B\to J/\psi \phi$ which can be
measured at the LHCb experiment.

$2-$ In supersymmetric theories, the effective Hamiltonian
$H_{\rm{eff}}^{\Delta B=2}$ receives new contributions through the
box diagrams mediated by gluino, chargino, neutralino, and charged
Higgs. It turns out that gluino exchanges give the dominant
contributions \cite{Ball:2003se}. The most general effective
Hamiltonian for $\Delta B=2$ processes, induced by gluino exchange
through $\Delta B=2$ box diagrams, can be expressed as \be
H^{\Delta B=2}_{\rm eff}=\sum_{i=1}^5 C_i(\mu) Q_i(\mu) +
\sum_{i=1}^3 \tilde{C}_i(\mu) \tilde{Q}_i(\mu) + h.c. \;,
\label{Heff} \ee where $C_i(\mu)$, $\tilde{C}_i(\mu)$ and
$Q_i(\mu)$, $\tilde{Q}_i(\mu)$ are the Wilson coefficients and the
local operators normalized at the scale $m_b$ respectively, which
can be found in Ref.\cite{Ball:2003se}. As in the $B_d$ system,
the effect of SUSY can be parameterized by a dimensionless
parameter $r_s$ and a phase $2\theta_s$ defined as follows:
\begin{equation}
r_s e^{i \theta_s} = \sqrt{\frac{{\cal M}_{12}(B_s)}{{\cal
M}^{\rm{SM}}_{12}(B_s)}}, \label{paramertization}
\end{equation}
where ${\cal M}_{12}(B_s) = \langle B^0_s \vert
H_{\mathrm{eff}}^{\Delta B=2} \vert \bar{B}^0_s \rangle \equiv
{\cal M}_{12}^{\rm{SM}} + {\cal M}_{12}^{\rm{SUSY}}$. Thus, the
total $B_s -\bar{B}_s$ mass difference is given by $\Delta M_{B_s}
= 2 \vert {\cal M}_{12}^{\rm{SM}}(B_s)\vert~ r_s^2=\Delta
M^{\rm{SM}}_{B_s}~ r_s^2$. In the mass insertion approximation,
the gluino contribution to the amplitude of $B_s$ oscillation is
given in terms of the ratio of the gluino mass to the average
squark mass, $x=m^2_{\tilde{g}}/m_{\tilde{q}}^2$, and the down
squark mass insertions between second and third generations,
$(\delta^d_{AB})_{23}$, where $A$ and $B$ stand for left $(L)$ or
right $(R)$ handed mixing. A general expression for $R_s= {\cal
M}^{\tilde{g}}_{12}/{\cal M}^{\rm{SM}}_{12} $ has been given in
Ref.\cite{Ball:2003se} as follows:
\begin{eqnarray}
R_s&=& a_1(m_{\tilde{q}},x) [(\delta^d_{LL})^2_{23} +
(\delta^d_{RR})^2_{23}] + a_2(m_{\tilde{q}},x)
[(\delta^d_{LR})^2_{23}\nonumber\\
&+& (\delta^d_{RL})^2_{23}] + a_3(m_{\tilde{q}},x)
[(\delta^d_{LR})_{23}(\delta^d_{RL})_{23}] \nonumber\\
&+& a_4(m_{\tilde{q}},x)
[(\delta^d_{LL})_{23}(\delta^d_{RR})_{23}]~, \label{rs}
\end{eqnarray}
where the coefficients $\vert a_1 \vert \simeq {\cal O}(1)$,
$\vert a_2 \vert < \vert a_3 \vert < \vert a_4 \vert \simeq {\cal
O}(100)$. For instance, with $m_{\tilde{q}}=300$ and $x=1$, one
finds
\begin{eqnarray}
R_s\!&\!=\!&\!7.2[(\delta^d_{LL})^2_{23} + (\delta^d_{RR})^2_{23}]
+ 129.8 [(\delta^d_{LR})^2_{23} +
(\delta^d_{RL})^2_{23}]\nonumber\\
\!&\!-\!&\!205.7 [(\delta^d_{LR})_{23}(\delta^d_{RL})_{23}] -
803.8 [(\delta^d_{LL})_{23}(\delta^d_{RR})_{23}].~
\end{eqnarray}
Note that $r_s^2 = \vert~ 1+R_s ~\vert$. From the experimental
upper bound on $\Delta M_{B_s}$ in Eq.(\ref{CDF}), one can derive
an upper bound on the mass insertions involved in Eq.(\ref{rs}).
In order to find conservative upper bounds, we set the SM
contribution to its best fit value, namely $\Delta
M^{\rm{SM}}_{B_s}=17.5~ \rm{ps}^{-1}$. In this case, the $\vert
R_s\vert$ should satisfy the following bound: \be \vert R_s \vert
=\left\vert \frac{(\Delta M_{B_s})_{Exp}}{(\Delta M_{B_s})_{SM}}
-1 \right\vert \lsim 4/17 .
\ee%
It is worth mentioning that if one assumes that $\Delta
M^{\rm{SM}}_{B_s}\simeq 21~ \rm{ps}^{-1}$, the above bound remains
valid. In table 1 we present our results for the upper bounds on
$\vert (\delta^d_{AB})_{23}\vert$ mass insertions from their
individual contributions to $B_s-\bar{B}_s$ mixing for
$m_{\tilde{q}}=300$ GeV and $x$ varies from $0.25$ to $2$. As can
be seen from Eq.(\ref{rs}) that the constraints imposed on the
mass insertions are symmetric under changing $L \leftrightarrow
R$. Therefore, we present in table 1 the upper bounds on one
combination of the mass insertions.

\begin{table}[t]
\begin{center}
{\small \begin{tabular}{|c|c|c|c|c|} \hline $x$ & $\vert
(\delta^d_{LL})_{23}\vert$ & $\vert (\delta^d_{LR})_{23}\vert$
&{\small $\sqrt{\vert
(\delta^d_{LR})_{23} (\delta^d_{RL})_{23} \vert}$}&{\small $\sqrt{\vert (\delta^d_{LL})_{23} (\delta^d_{RR})_{23}\vert}$} \\
\hline \hline
0.25 & 0.074 & 0.035 & 0.018 & 0.014 \\
\hline
 0.5 & 0.11 & 0.037 & 0.024 & 0.015 \\
\hline
 1 & 0.17 & 0.04 & 0.032 & 0.016 \\
\hline
 1.5 & 0.27 & 0.43 & 0.039 & 0.017 \\
\hline
 2 & 0.46 & 0.046 & 0.046 & 0.018 \\
 \hline
\end{tabular}}
\caption{Upper bounds on $\vert(\delta^d_{AB})_{23}\vert$,
$\{A,B\}=\{L,R\}$ from $\Delta M_{B_s} < 21~ \rm{ps}^{-1}$ for
$m_{\tilde{q}}=300~\rm{GeV}$.}
\end{center}
\end{table}

Three comments on the results of table 1 are in order: $1)$ the
constraints obtained on $\vert (\delta^d_{LL(RR)})_{23}\vert$ are
the strongest known constraints on these mass insertions, since
other processes based on $b\to s$ transition, like $B\to X_s
\gamma$, leave them unconstrained \cite{Khalil:2005qg}. In fact
with these constraints, one can verify that the $LL(RR)$
contributions to $B\to \phi K$, $B\to \eta' K$ and $B\to \pi K$
are diminished and become insignificant. Therefore, $LR$
contribution remains as the only candidate for saturating any
deviation from the SM results in the CP asymmetries or branching
ratios of these processes \cite{Khalil:2005dx}. $2)$ The upper
bounds on $LR(RL)$ mass insertions from the $B_s -\bar{B}_s$ are
less stringent than those derived from the experimental limits of
the branching ratio of $B\to X_s \gamma$ \cite{Khalil:2005qg}.
$3)$ The combined effect of $(\delta^d_{LL})_{23}$ and
$(\delta^d_{RR})_{23}$ is severely constrained by $\Delta
M_{B_s}$. However, the lowest value of $(\delta^d_{LL})_{23}$,
that can be obtained in the minimal SUSY model with universal soft
SUSY breaking terms, is of order $\lambda^2 \sim {\cal
O}(10^{-2})$. Therefore, it is clear that models with large $RR$
mixing would be disfavored by the $\Delta M_{B_s}$ constraints,
consistently with the previous conclusions reached by using the
mercury electric dipole moment (EDM) constraints
\cite{Abel:2004te}. Indeed, with a large $(\delta^d_{RR})_{23}$
one may induce a large imaginary part of the mass insertion
$(\delta^d_{LR})_{22}$ which overproduces the mercury EDM. This
also implies strong constraints on the right squark mixings.

$3-$ The $B_s\to J/\psi\phi$ decay is accessible at hadron
colliders where plenty of $B_s$ will be produced. It is,
therefore, considered as one of the benchmark channels to be
studied at LHCb experiment. The final state of $B_s\to J/\psi
\phi$ is not a CP eigenstate, but a superposition of CP odd and
even states which can, however, be disentangled through an angular
analysis of their products \cite{Dighe:1995pd}. This angular
distribution yields to a tiny direct CP violation. Thus, the CP
asymmetry of the $B_s$ and $\bar{B}_s$ meson decay to $J/\psi
\phi$ is given by \bea a_{J/\psi \phi}(t) &=&
\frac{\Gamma(\bar{B}^0_s(t) \to J/\psi \phi) - \Gamma(B^0_s(t) \to
J/\psi \phi)}{\Gamma(\bar{B}^0_s(t) \to J/\psi \phi) +
\Gamma(B^0_s(t) \to J/\psi \phi)}\nonumber\\
&=& S_{J/\psi \phi} \sin(\Delta M_{B_s} t), \eea where $S_{J/\psi
\phi}$ is the mixing-induced  CP asymmetry. In the SM, the mixing
CP asymmetry $S_{J/\psi \phi}$ is given by \cite{Ball:2003se} \bea
S_{J/\psi \phi} &=& \sin 2 \beta_s^{\rm{SM}}= \sin\left[2~
\rm{arg}(V_{tb} V_{ts}^*)\right]\nonumber\\
& \simeq& -2 \lambda^2 \eta \simeq {\cal O}(10^{-2}).\eea Such
small CP asymmetry in the SM gives the hope that if a sizable
value of $S_{J/\psi\phi}$ is found in future experiments (in
particular at LHCb experiment), then it would be an immediate
signal for a new physics effect.

In the presence of SUSY contribution, the CP asymmetry $S_{J/\psi
\phi}$ is given by \cite{Ball:2003se} \be S_{J/\psi\phi} = \sin 2
\beta^{\rm{eff}}_{s} = \sin \left(2\beta^{\rm{SM}}_s + 2 \theta_s
\right), \ee where $\theta_s$ is given in
Eq.(\ref{paramertization}) as $2 \theta_s =
\rm{arg}\left(1+R_s\right)$. Therefore, the value of
$S_{J/\psi\phi}$ depends on the magnitude of $R_s$ which, as
emphasized above, is constrained from $\Delta M_{B_s}$ to be less
than or equal $4/17$. In this respect, it is easy to show that the
maximum value of $S_{J/\psi\phi}$ that one may obtain from SUSY
contributions to the $B_s-\bar{B}_s$ mixing is given by \be
S_{J/\psi\phi}\simeq 0.24. \ee It is important to note that due to
the stringent constraints on $(\delta^d_{LR})_{23}$ from $b\to
s\gamma$: $\vert (\delta^d_{LR})_{23} \vert \lsim 0.016$, the LR
(RL) supersymmetric contribution to $S_{J/\psi\phi}$ is very
restricted. It implies that $S_{J/\psi\phi} < 0.02$, which is too
small to be observed at the Tevatron or the LHC. Therefore, the LR
and RL contributions can not provide significant contribution to
$B_s$ mixing or to the mixing CP asymmetry of $B_s \to J/\psi
\phi$.

On the other hand, the $LL$ and $RR$ mass insertions can generate
sizable and measurable values of $S_{J/\psi\phi}$. For instance,
$(\delta^d_{LL(RR)})_{23} \simeq 0.17~ e^{i \pi/4}$ yields to $R
\simeq 0.24~ e^{i\pi/2}$ which implies that $\sin 2\beta_s \simeq
0.24$. However, as mentioned above, it is important to note that
since the minimum value of the mass insertion
$(\delta^d_{LL})_{23}$ is of order $10^{-2}$, thus, in case of
SUSY models with large right-handed squark mixings, \ie,
$(\delta^d_{RR})_{23}\sim 0.17$, one finds that
$\sqrt{(\delta^d_{LL})_{23}(\delta^d_{RR})_{23}} \sim {\cal
O}(10^{-1})$ which may exceed its upper bound presented in table
1. Therefore, in this scenario, contributions from both
$(\delta^d_{RR})_{23}^2$ and
$(\delta^d_{LL})_{23}(\delta^d_{RR})_{23}$ should be considered
simultaneously in determining the $\Delta M_{B_s}$ and $\sin
2\beta_s$.

$4-$ We now consider the impact of the $\Delta M_{B_s}$
constraints derived above on the mixed CP asymmetries in $B_d \to
\phi K$ and $B_d \to \eta'K$ processes, which at the quark level
are also based on $b\to s$ transition. The BaBar and Belle results
for these asymmetries lead to the following averages: \be S_{\phi
K} = 0.47 \pm 0.19,~~~~~~~~~~~~~~~~ S_{\eta' K} = 0.48 \pm 0.09,
\label{average}\ee which display about $1\sigma$ and $2.5 \sigma$
deviation from the SM predictions, respectively.

The SUSY contributions to the decay amplitudes of $B_d\to \phi K$
and $B_d\to \eta' K$ are given by \cite{Gabrielli:2005ys} \bea
A_{\phi K}\!&\!=\!&\! - i \frac{G_F}{\sqrt{2}} m_{B_d}^2
F_+^{B_d\to K} f_{\phi} \sum_{i=1}^{12} H_{i}(\phi) (C_{i} +
\tilde{C}_{i}), \nonumber\\
A_{\eta' K}\!&\!=\!&\! - i \frac{G_F}{\sqrt{2}} m_{B_d}^2
F_+^{B_d\to K} f_{\eta'}^s \sum_{i=1}^{12} H_{i}(\eta') (C_{i} -
\tilde{C}_{i}),~~~~ \label{APhi} \eea where the $C_i$ are the
corresponding Wilson coefficients to the local operators of $b\to
s$ transition. $C_i$ as functions of the mass insertions
$(\delta^d_{LL})_{23}$ and $(\delta^d_{LR})_{23}$ and
$\tilde{C}_i$ as functions of $(\delta^d_{RR})_{23}$ and
$(\delta^d_{RL})_{23}$ can be found in
Ref.\cite{Gabrielli:2005ys}. Here the QCD factorization mechanism
is adopted to determine the hadronic matrix elements and as in
Ref.\cite{Gabrielli:2005ys} they can be parameterized in terms of
the parameters $H_i(\phi)$ and $H_i(\eta')$ which are given in
Ref.\cite{Gabrielli:2005ys}. In terms of SUSY contributions, the
CP asymmetry $S_{\phi(\eta')K}$ can be written as
\cite{Khalil:2002fm} \be S_{\phi(\eta')K} = \sin 2\beta + 2 \vert
R_{\phi(\eta')}\vert \cos \delta_{\phi(\eta')} \sin
\theta_{\phi(\eta')} \cos 2 \beta,~~~~~ \ee where
$R_{\phi(\eta')}=
\left(\frac{A^{\rm{SUSY}}}{A^{\rm{SM}}}\right)_{\phi(\eta') K}$~,~
$\theta_{\phi(\eta')}=\rm{arg} \left[
\left(\frac{A^{\rm{SUSY}}}{A^{\rm{SM}}}\right)_{\phi(\eta') K}
\right]$ and $\delta_{\phi(\eta')}$ is the strong phase. Thus, in
order to derive $S_{\phi(\eta')K}$ toward their central values of
the average experimental results in Eq.(\ref{average}), $\vert
R_{\phi(\eta')}\vert \gsim 0.2$ should be satisfied. For a gluino
mass and average squark mass of order
$\tilde{m}=m_{\tilde{g}}=500$ GeV, one finds \be R_{\phi}= -\,
0.14 \, e^{-i\,0.1} (\delta^d_{LL})_{23}\,-\, 127\, e^{-i\,0.08}
(\delta^d_{LR})_{23} \,+\, L\leftrightarrow R, \ee and \be
R_{\eta'}= -\, 0.07 \, e^{i\,0.24} (\delta^d_{LL})_{23}\,-\, 64\,
(\delta^d_{LR})_{23} \,-\, L\leftrightarrow R. \ee It is now clear
that the $\Delta M_{B_s}$ constraints play a crucial role in
reducing the $LL$ and $RR$ contributions to the
$S_{\phi(\eta')K}$. By implementing the bounds in table 1, one can
easily observe that $LL(RR)$ contribution leads to $\vert
R_{\phi(\eta')} \vert \sim {\cal O}(10^{-2})$ which yields a
negligible effect on $S_{\phi(\eta')}$ and one can safely conclude
that the $LL$ and $RR$ mass insertions can not provide an
explanation to any deviation in $S_{\phi(\eta')}$ results. On the
other hand, the contribution of $(\delta^d_{LR})_{32}$ is less
constrained by $\Delta M_{B_s}$ and large effects in $\vert
R_{\phi(\eta')} \vert$ that could drive $S_{\phi(\eta') K}$ toward
$0.4$ can be achieved.

$5-$ The above results show that $S_{J/\psi \phi}$ and
$S_{\phi(\eta') K}$ are dominated by different mass insertions:
$LL$ and $LR/RL$ respectively. As emphasized in
Ref.\cite{Khalil:2005dx}, these two mass insertions can be
enhanced simultaneously in SUSY models with intermediate/large
$\tan\beta$ and a simple non-minimal flavor structure, where the
scalar mass of the first two generations is different from the
scalar mass of the third generation. In particular, let us
consider the following soft SUSY breaking terms are assumed at the
GUT scale. \bea &&M_1=M_2 =M_3 = M_{1/2},~~~~  A^u = A^d =
A_0 e^{i \phi_A},\nonumber\\
&&M_{U}^2 = M_{D}^2 = m_0^2,~~~~~~~~~~~~~~m_{H_1}^2 = m_{H_2}^2 =
m_0^2,\nonumber\\ &&M_{Q}^2 = \left( \matrix{ m_0^2 & & \cr
 & m_0^2 &  \cr
 &  & a^2 m_0^2 }\right) \;.
\eea The parameter $a$ measures the non-universality of the squark
masses. It is worth mentioning that the EDM constraints on the CP
violating phase $\phi_A$ of the trilinear coupling is less sever
than the constraints imposed on the other SUSY CP phases and can
be of order ${\cal O}(0.1)$ \cite{Abel:2001vy}.

Using the relevant renormalization group equations, one can
explore these parameters from GUT scale to the electroweak scale,
where we impose the electroweak symmetry breaking conditions and
calculate the squark mass matrices. Then we determine the
numerical values of the corresponding mass insertions. For
instance, for $a=5$, $\tan \beta =15$ and $m_{\tilde{g}}\sim
m_{\tilde{q}} \sim 500$ GeV, one finds that $
\vert(\delta^d_{LL})_{23} \vert \simeq 0.18 $ which leads to
$\Delta M_{B_s} \simeq 19~ \rm{ps}^{-1}$. Also with a proper
choice for the phase $\phi_A$, one can get
$\rm{arg}[(\delta^d_{LL})_{23}] \simeq 0.7$ which implies that $
S_{J/\psi \phi} \simeq 0.1$ which can be measured by the LHCb
experiment. Note that in this scenario the phases of the mass
insertions are due to a combined effect of the SM phase in the CKM
mixing matrix and the SUSY CP phase $\phi_A$. However, for the
$LL$ mass insertion the main effect is due the CKM phase, see
Ref.\cite{Khalil:2005dx} for more details.

Concerning the mass insertion $(\delta^d_{LR})_{23}$, it is
expected to be negligible due to the universality of the trilinear
couplings. However, with intermediate/large $\tan \beta$, the
double mass insertion is quite important and it gives the dominant
effect as follows \cite{Khalil:2005dx} \bea
(\delta^d_{LR})_{23_{\rm{eff}}} = (\delta^d_{LR})_{23} +
(\delta^d_{LL})_{23}~ (\delta^d_{LR})_{33}, \eea where
$(\delta^d_{LR})_{33}\simeq \frac{m_b (A_b - \mu \tan
\beta)}{\tilde{m}^2}$. Since $(\delta^d_{LR})_{23}$ is negligible,
$(\delta^d_{LR})_{23_{\rm{eff}}}$ is given by \bea
(\delta^d_{LR})_{23_{\rm{eff}}} \simeq (\delta^d_{LL})_{23}~
\frac{m_b}{\tilde{m}}~ \tan\beta.\eea The parameter $\mu$ is
determined by the electroweak conditions and it is found to be of
order the squark mass. The phase of $\mu$ set to zero to overcome
the EDM constraints. Since $(\delta^d_{LL})_{23} \simeq 0.18$, the
value of $(\delta^d_{LR})_{23_{\rm{eff}}}$ is of order $10^{-2}$
which is sufficient to reduce the CP asymmetries $S_{\phi K}$ and
$S_{\eta' K}$ from the SM result $\sin 2\beta\simeq 0.7$ to their
central values of average experimental results.

$6-$ To conclude, We have considered the supersymmetric
contributions to the $B_s -\bar{B}_s$ mixing. We derived new model
independent constraints on the magnitude of the mass insertions
$(\delta^d_{AB})_{23}$, where $\{A,B\}=\{L,R\}$, from the new
experimental measurements of $\Delta M_{B_s}$. We showed that by
implementing these constraint, the SUSY contribution, through the
$LL$ mixing, can enhance the CP asymmetry of $B_s \to J/\psi \phi$
up to $0.24$, which can be observed at the LHCb experiment. We
also emphasized that the new constraints exclude the SUSY models
with large RR flavor mixing and severely restrict the $LL$
contributions to the CP asymmetries of $B\to \phi K$ and $B \to
\eta' K$. Therefore, SUSY models with dominant LR flavor mixing is
the only way to accommodate the apparent deviation of CP
asymmetries from those expected in the standard model. Finally we
studied an example of SUSY non-minimal flavor model and
intermediate/large $\tan\beta$. We showed that in this model the
new $\Delta M_{B_s}$ results and also the CP asymmetries in $B_s
\to J/\psi \phi$, $B\to \phi K$ and $B \to \eta' K$ processes can
be simultaneously saturated.

%%%%%%%%%%%%%%%%%%%%%%%%%%%%%%%%%%%%%%%%%%%

\end{document}